\documentclass[twocolumn]{aastex631}

\usepackage{upgreek}
\usepackage{textcomp}
\usepackage{wasysym}
\usepackage{blindtext}
\usepackage{lineno}
\usepackage{rotating}
\usepackage{hyperref}
\usepackage{subfigure}
\usepackage{longtable}
\usepackage{tablefootnote}
\usepackage{appendix}
\usepackage{CJKutf8}
\usepackage{afterpage}
\usepackage{booktabs}
\usepackage{array} 
\usepackage{float}
\usepackage{etoolbox}
\usepackage[T1]{fontenc}
\usepackage{lmodern}
\usepackage{silence}
\shorttitle{New \texttt{SMDET} Discoveries}
\shortauthors{Brooks et al.}

\begin{document}

\title{Discovery of 118 New Ultracool Dwarf Candidates Using Machine Learning Techniques}

\author[0000-0002-5253-0383]{Hunter Brooks}
\affiliation{Department of Astronomy and Planetary Science, Northern Arizona University, Flagstaff, AZ 86011, USA}
\affiliation{NSF National Optical-Infrared Astronomy Research Laboratory, 950 N. Cherry Ave., Tucson, AZ 85719, USA}

\author[0000-0001-7896-5791]{Dan Caselden}
\affiliation{Department of Astrophysics, American Museum of Natural History, Central Park West at 79th Street, New York, NY 10024, USA}

\author[0000-0003-4269-260X]{J.\ Davy Kirkpatrick}
\affiliation{IPAC, Mail Code 100-22, Caltech, 1200 E. California Blvd., Pasadena, CA 91125, USA}

\author[0000-0001-9778-7054]{Yadukrishna Raghu}
\affiliation{Backyard Worlds: Planet 9}

\author{Charles A. Elachi}
\affiliation{Seaver College, Pepperdine University, Malibu, CA 90263, USA}

\author[0000-0002-2466-865X]{Jake Grigorian}
\affiliation{University of Southern California, University Park Campus, Los Angeles, CA 90089, USA}

\author[0009-0008-3778-487X]{Asa Trek}
\affiliation{Backyard Worlds: Planet 9}

\author[0009-0005-6222-6026]{Andrew Washburn}
\affiliation{Backyard Worlds: Planet 9}

\author[0009-0004-9088-7510]{\begin{CJK*}{UTF8}{} Hiro Higashimura ({\CJKfamily{min}東村滉}) \end{CJK*}}
\affiliation{Earl of March Intermediate School, 4 The Pkwy, Kanata, ON K2K 1Y4, Canada}

\author[0000-0002-1125-7384]{Aaron M. Meisner}
\affiliation{NSF National Optical-Infrared Astronomy Research Laboratory, 950 N. Cherry Ave., Tucson, AZ 85719, USA}

\author[0000-0002-6294-5937]{Adam C. Schneider}
\affiliation{United States Naval Observatory, Flagstaff Station, 10391 West Naval Observatory Rd., Flagstaff, AZ 86005, USA} 

\author[0000-0001-6251-0573]{Jacqueline K. Faherty}
\affiliation{Department of Astrophysics, American Museum of Natural History, Central Park West at 79th Street, New York, NY 10024, USA}

\author[0000-0001-7519-1700]{Federico Marocco}
\affiliation{IPAC, Mail Code 100-22, Caltech, 1200 E. California Blvd., Pasadena, CA 91125, USA}

\author{Christopher R. Gelino}
\affiliation{IPAC, Mail Code 100-22, Caltech, 1200 E. California Blvd., Pasadena, CA 91125, USA}

\author[0000-0002-2592-9612]{Jonathan Gagn\'e}
\affiliation{Plan\'etarium de Montr\'al, Espace pour la Vie, 4801 av. Pierre-de Coubertin, Montr\'eal, Qu\'ebec, Canada}
\affiliation{Trottier Institute for Research on Exoplanets, Universit\'e de Montr\'eal, D\'epartement de Physique, C.P.~6128 Succ. Centre-ville, Montr\'eal, QC H3C~3J7, Canada}

\author[0000-0003-2235-761X]{Thomas P. Bickle}
\affil {School of Physical Sciences, The Open University, Milton Keynes, MK7 6AA, UK}
\affil{Backyard Worlds: Planet 9}

\author[0000-0003-4247-1401]{Shih-Yun Tang}
\affiliation{Department of Physics and Astronomy, Rice University, 6100 Main Street, Houston, TX 77005, USA}
\affiliation{Lowell Observatory, 1400 West Mars Hill Road, Flagstaff, AZ 86001, USA}

\author[0000-0003-4083-9962]{Austin Rothermich}
\affiliation{Department of Astrophysics, American Museum of Natural History, Central Park West at 79th Street, New York, NY 10024, USA}
\affiliation{Department of Physics, Graduate Center, City University of New York, 365 5th Avenue, New York, NY 10016, USA}
\affiliation{Department of Physics and Astronomy, Hunter College, City University of New York, 695 Park Avenue, New York, NY 10065, USA; Backyard Worlds: Planet 9, USA}

\author[0000-0002-6523-9536]{Adam J.\ Burgasser}
\affiliation{Department of Astronomy \& Astrophysics, UC San Diego, La Jolla, CA, USA}

\author[0000-0002-2387-5489]{Marc J. Kuchner}
\affiliation{NASA Goddard Space Flight Center, Exoplanets and Stellar Astrophysics Laboratory, Code 667, Greenbelt, MD 20771, USA}

\author{Paul Beaulieu}
\affiliation{Backyard Worlds: Planet 9}

\author{John Bell}
\affil{Backyard Worlds: Planet 9}

\author[0000-0002-7630-1243]{Guillaume Colin}
\affiliation{Backyard Worlds: Planet 9}

\author[0000-0002-8295-542X]{Giovanni Colombo}
\affiliation{Backyard Worlds: Planet 9}

\author{Alexandru Dereveanco}
\affiliation{Backyard Worlds: Planet 9}

\author{Deiby Pozo Flores}
\affiliation{Backyard Worlds: Planet 9}

\author{Konstantin Glebov}
\affiliation{Backyard Worlds: Planet 9}

\author[0000-0002-8960-4964]{Leopold Gramaize}
\affiliation{Backyard Worlds: Planet 9}

\author[0000-0002-7389-2092]{Les Hamlet}
\affiliation{Backyard Worlds: Planet 9}

\author[0000-0002-4733-4927]{Ken Hinckley}
\affiliation{Backyard Worlds: Planet 9}

\author[0000-0003-4905-1370]{Martin Kabatnik}
\affiliation{Backyard Worlds: Planet 9}

\author[0000-0001-8662-1622]{Frank Kiwy}
\affiliation{Backyard Worlds: Planet 9}

\author{David W. Martin}
\affiliation{Backyard Worlds: Planet 9}

\author{Raúl F. Palma Méndez}
\affiliation{Backyard Worlds: Planet 9}

\author{Billy Pendrill}
\affiliation{Backyard Worlds: Planet 9}

\author{Lizzeth Ruiz}
\affiliation{Backyard Worlds: Planet 9}

\author{John Sanchez}
\affiliation{Backyard Worlds: Planet 9}

\author[0000-0003-4864-5484]{Arttu Sainio}
\affiliation{Backyard Worlds: Planet 9}

\author[0000-0002-7587-7195]{J\"{o}rg Sch\"{u}mann}
\affiliation{Backyard Worlds: Planet 9}

\author{Manfred Schonau}
\affiliation{Backyard Worlds: Planet 9}

\author{Christopher Tanner}
\affiliation{Backyard Worlds: Planet 9}

\author[0000-0003-4714-3829]{Nikolaj Stevnbak}
\affiliation{Backyard Worlds: Planet 9}

\author{Andres Stenner}
\affiliation{Backyard Worlds: Planet 9}

\author[0000-0001-5284-9231]{Melina Thévenot}
\affiliation{Backyard Worlds: Planet 9}

\author{Vinod Thakur}
\affiliation{Backyard Worlds: Planet 9}

\author{Nikita V. Voloshin}
\affiliation{Backyard Worlds: Planet 9}

\author{Zbigniew W\c{e}dracki}
\affiliation{Backyard Worlds: Planet 9}

\collaboration{100}{The Backyard Worlds: Planet 9 Collaboration}
\affiliation{}

\begin{abstract}
We present the discovery of 118 new ultracool dwarf candidates, discovered using a new machine learning tool, named \texttt{SMDET}, applied to time series images from the {\it Wide-field Infrared Survey Explorer}. We gathered photometric and astrometric data to estimate each candidate's spectral type, distance, and tangential velocity. This sample has a photometrically estimated spectral class distribution of 28 M dwarfs, 64 L dwarfs, and 18 T dwarfs. We also identify a T subdwarf candidate, two extreme T subdwarf candidates, and two candidate young ultracool dwarfs. Five objects did not have enough photometric data for any estimations to be made. To validate our estimated spectral types, spectra were collected for 2 objects, yielding confirmed spectral types of T5 (estimated T5) and T3 (estimated T4). \texttt{SMDET} demonstrates the effectiveness of machine learning tools as a new large-scale discovery technique. 
\end{abstract}


\section{Introduction}
The continued discovery of new brown dwarf candidates enables a better understanding of the initial mass function at low masses, and modeling of the formation and evolution of local stellar populations (\citealt{Dantona Mazzitelli(1986)}). With recent discoveries and advancements in the study of exoplanets, it has become increasingly clear that brown dwarfs and large gaseous exoplanets have similar characteristics (\citealt{Faherty et al.(2021)}). Therefore, increasing the variety and number of identified brown dwarfs results in a more diverse set of analogs against which to compare exoplanets. 

One method to discover large numbers of brown dwarfs is to survey catalogs for objects with colors similar to those of known brown dwarfs. For example, \cite{Kirkpatrick et al.(2011)} employed this technique using early data from the \textit{Wide-field Infrared Survey Explorer} (\textit{WISE}; \citealt{Wright et al. 2010}), focusing on the W1$-$W2 color to identify potential brown dwarfs and applying a criterion to remove extragalactic sources: W1 $-$ W2 \textgreater 0.96(W2 $-$ W3) $-$ 0.96. However, purely imaging-based discovery can still lead to sample contamination from extragalactic sources, even with removal criteria in place.

Another approach to brown dwarf discovery is to survey based on proper motion, which helps exclude extragalactic sources as contaminants. Several proper motion surveys have been conducted across various telescopes, an example includes using \textit{WISE} imaging due to the long time baseline of the \textit{WISE/NEOWISE} mission (\citealt{Kirkpatrick et al.(2014)}, \citealt{Luhman(2014)}, \citealt{Kirkpatrick et al. 2016}, \citealt{Schneider et al.(2016)}, \citealt{Kuchner et al. 2017}, \citealt{Greco et al.(2020)}, \citealt{Kota et al.(2022)}, and others). However, a search strategy that focuses on proper motion may miss objects whose space motions are primarily along the line of sight. Moreover, for brown dwarfs at further distances detecting its proper motions will be increasingly difficult unless the time baseline or motion is large.

This paper applies a novel proper motion detection technique to discover a large sample of faint, fast-moving objects, using a new convolutional neural network model called \texttt{SMDET} was developed (\citealt{Caselden et al.(2020)} and \textcolor{blue}{Caselden et al. 2024, in preparation}). \texttt{SMDET}'s technique and the new candidate sample are described in \S\ref{new}. Photometry and astrometry for each object were gathered as described in \S\ref{photometry}. \S\ref{spectral} discusses photometric spectral type estimation, for which we use the photo-type method discussed in \cite{Skrzypek et al.(2015),Skrzypek et al.(2016)} . With these photometric spectral types, we estimate distances and tangential velocities in \S\ref{distance}. A handful of objects, not following the general color trends laid out in \cite{Kirkpatrick et al.(2021)} for M, L, T, and Y dwarfs, found to be candidate subdwarfs as discussed in \S\ref{cheetoo}. We discuss the identification of possible young ultracool dwarfs in our study in \S\ref{young}, and the limitations of our study in \S\ref{dis}. Our main results are summarized in \S\ref{conclusion}.

\section{Methods For Discovering New Ultracool Dwarf Candidates} \label{new}
\subsection{\texttt{SMDET}}
Brown dwarfs are challenging to detect because they are inherently dim and emit primarily at infrared wavelengths. Many known brown dwarfs were discovered using infrared imaging, including \textit{WISE} data (e.g., \citealt{Cushing et al.(2011)}, \citealt{Kuchner et al. 2017},  \citealt{Meisner et al.(2020a)}, \citealt{Kirkpatrick et al.(2021)}). These brown dwarfs are generally nearby and, as a result, predominantly exhibit high proper motions. However, it has proven relatively challenging to discover high proper motion brown dwarfs ($\geq$ 0.35 arcsec yr$^{-1}$) significantly fainter than the WISE W2 single-exposure detection limit of W2 = 14.5 Vega mag (\citealt{Faherty et al.(2009)} and \citealt{Bihain Scholz(2016)}). This limitation motivated us to search for such brown dwarfs using \texttt{SMDET} (\citealt{Caselden et al.(2020)} and \textcolor{blue}{Caselden et al. 2024, in preparation}), a neural network that detects faint, fast-moving objects in \textit{WISE} images.

\begin{figure*}[ht]
        \centering
        \includegraphics[scale = 0.9]{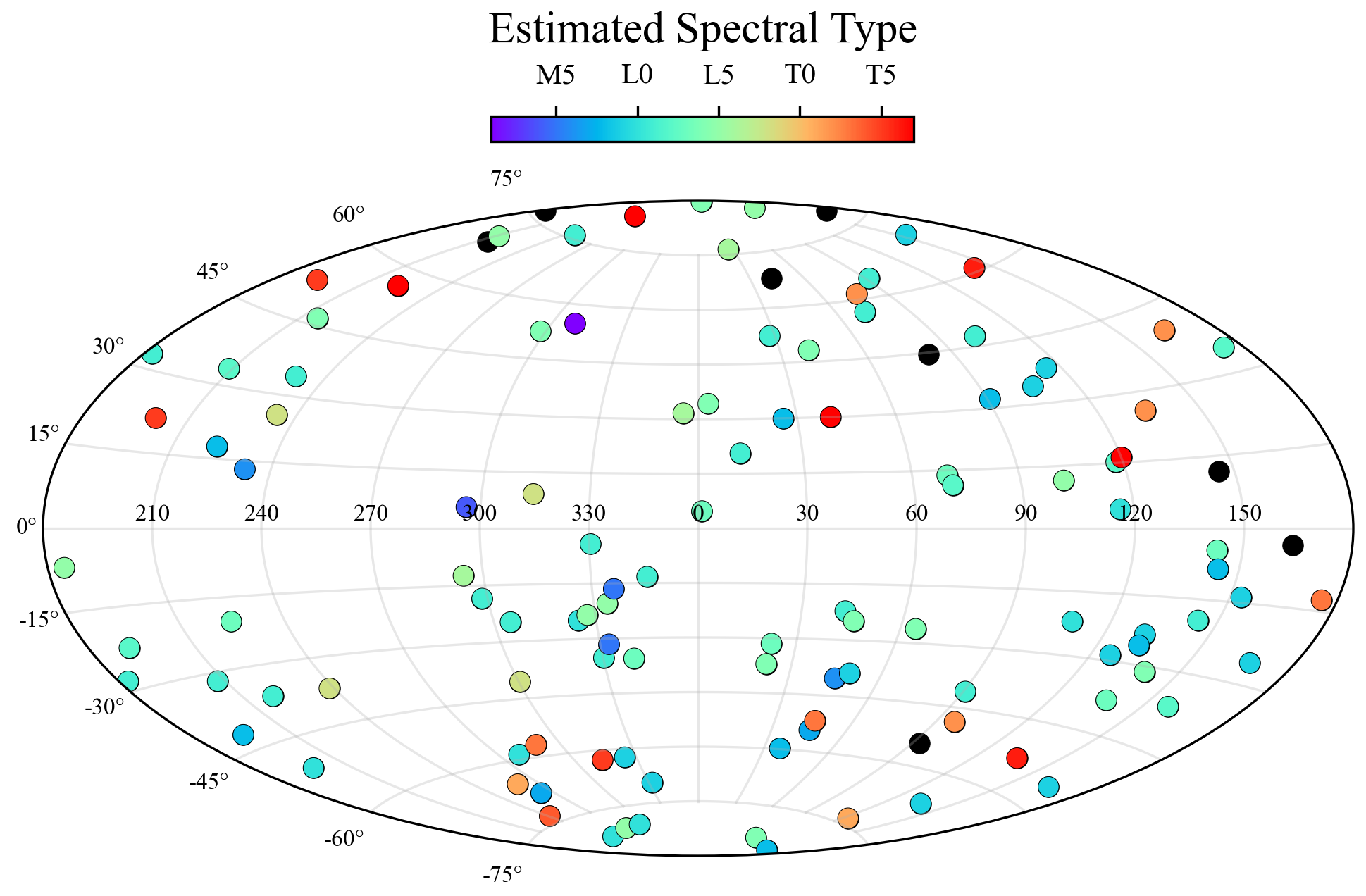}
        \caption{All-sky distribution of the 118 new \texttt{SMDET} discoveries in equatorial coordinates with the Aitoff projection. Sources are color-coded based on spectral type estimates from \S\ref{spectral}, while those with a black interior are either candidate subdwarfs discussed in \S\ref{cheetoo} or lack a photometric spectral type estimate.}
        \label{Figure X}
\end{figure*}

The unWISE project reprocessed \textit{WISE} exposures into coadds preserving the native \textit{WISE} angular resolution (\citealt{Meisner et al.(2018)}). \texttt{SMDET} was trained with synthetic objects added to unWISE time series coadds. Preliminary sky region rankings and segmented time series images were produced by \texttt{SMDET} across the entire sky with unWISE time series coadds. Sky regions were ranked according to the presence of faint, fast objects. The segmented time series images indentify which pixels in the input capture 1$\%$ or more of a faint, fast object's flux. For additional information regarding the backbone of \texttt{SMDET} see \citealt{Caselden et al.(2020)} and \textcolor{blue}{Caselden et al. 2024, in preparation}.

\subsection{Human Verification}\label{human}

After the training of \texttt{SMDET}, candidates drawn from regions with the highest scores were given to citizen scientist collaborators to visually scrutinize for the presence of real motion objects, distinguishing them from artifacts like diffraction spikes and ghosts. During the process of visually vetting \texttt{SMDET} candidates, 1,730 real proper motion objects and 10,170 \textit{WISE} artifacts, diffraction spikes and ghosts, were identified out of 11,900 \texttt{SMDET} candidates inspected by our team. 

\subsection{Filtering Out Known Objects}

We searched the SIMBAD Astronomical Database \cite{Wenger et al.(2000)} and the VizieR Catalog Access Tool \citep{Cutri et al.(2021)} to determine which of the 1,730 confirmed proper motion objects might be previously cataloged. We used a search radius of 1.0$\arcmin$ and followed up with visual confirmation of the cross-matched object with our real motion objects, resulting in 232 previously uncataloged sources.

After exploring the literature, we searched Gaia Data Release 3 (hereafter, Gaia DR3; \citealt{Gaia Collaboration et al.(2023)}) to eliminate objects that matched Gaia sources, as our focus is on cold brown dwarfs that are unlikely to be detectable with Gaia. We used the Wide-field Retrieval of Astrodata Program (hereafter, WRAP; \citealt{Brooks et al.(2023)}) Gaia search feature with a radius of 150$\arcsec$. The large search radius was necessary 
to confirm objects that are in crowded fields. This left 118 new high proper motion ultracool dwarf candidates. These candidates are presented in Table \ref{Table 2} and their sky distribution is illustrated in Figure \ref{Figure X}. The plot indicates that there is no apparent bias towards either the galactic plane or the poles. Note that many of these candidates were independently discovered as moving object candidates through the Backyard Worlds: Planet 9 citizen science project (\citealt{Kuchner et al. 2017}). Volunteer citizen scientists who also identified these objects are included in the ‘Discoverers’ column in Table \ref{Table 2}.

\begin{figure*}[ht]
        \centering
        \includegraphics[scale = 0.4]{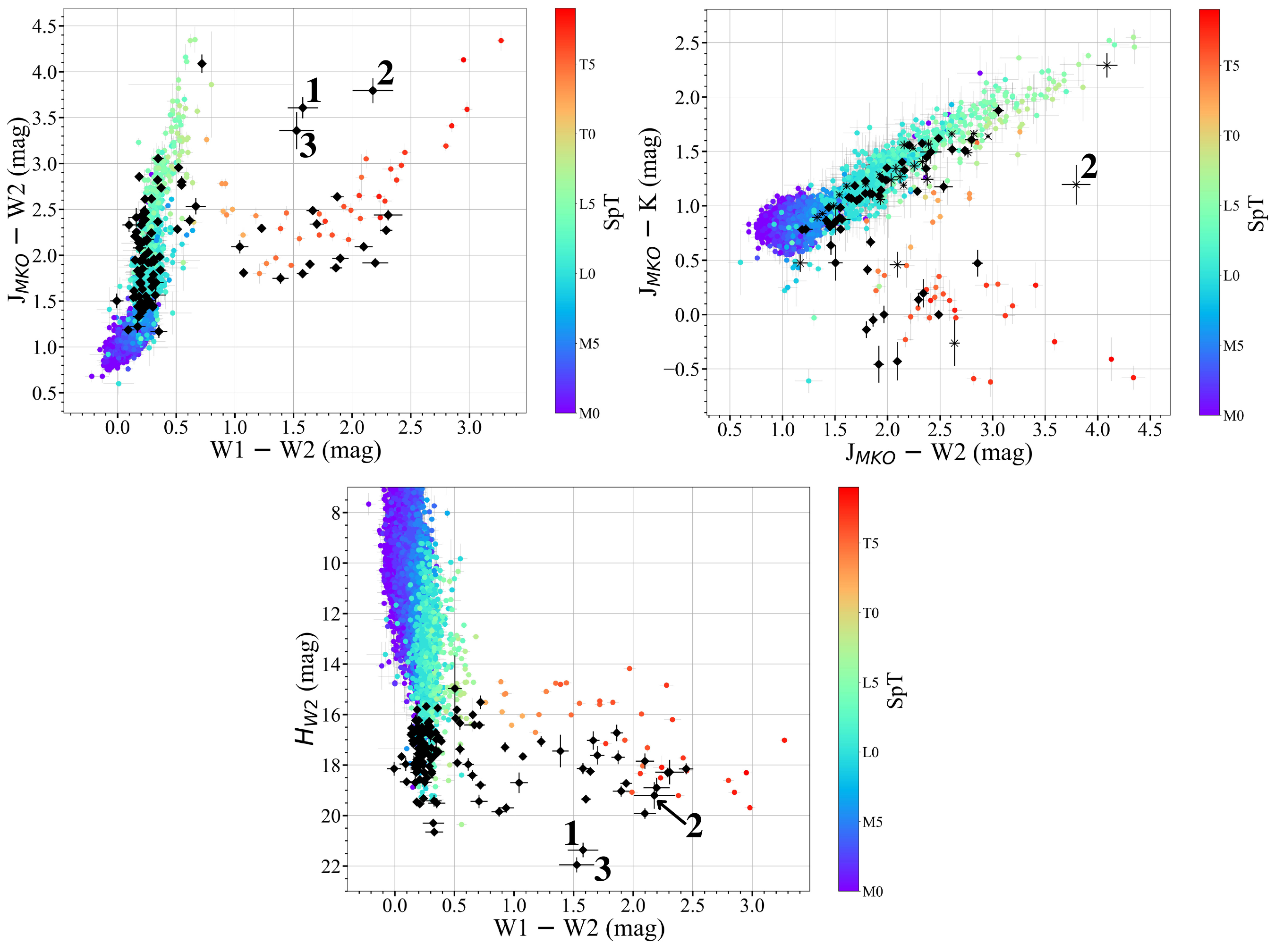}
        \caption{Color-color and reduced magnitude-color diagrams for
        \texttt{SMDET} discoveries (black points) and known ultracool dwarfs (colored data points) from the PanSTARRS 3$\pi$ survey (\citealt{Best et al.(2018)}). Objects from VHS and observed with the $K_{s}$ filter are marked as X symbols. Objects from UKIDSS LAS DR9 and observed with the $K_{MKO}$ filter are marked as diamond symbols. Points associated with numbers refer to subdwarf candidates discussed in \S\ref{cheetoo}, and correspond to 1: CWISE J024822.37+674812.6, 2: CWISE J052544.47+443409.9, and 3: CWISE J120642.54+762641.6}
        \label{Figure 2}
\end{figure*}

\section{Photometry and Astrometry} \label{photometry}

Archival photometric data were collected for the 118 discoveries to help with source characterization. The catalogs used to gather the photometry were CatWISE2020 (\citealt{Marocco et al. 2021}), the $AllWISE$ Source Catalog (hereafter, $AllWISE$; \citealt{Cutri et al.(2021)}), the VISTA Hemisphere Survey \footnote{Only covers southern hemisphere, VHS K filter: K$_s$} (hereafter, VHS; \citealt{Cross et al. (2012)}), the UKIRT Infrared Deep Sky Survey Large Area Survey Data Release 9/UKIRT Hemisphere Survey \footnote{Only covers northern hemisphere, UKIDSS LAS DR9 K filter: K$_{MKO}$} (hereafter, UKIDSS LAS DR9/UHS; \citealt{Lawrence et al.(2007)} and \citealt{Dye et al.(2018)}), and PanSTARRS Data Release 2 (hereafter, PanSTARRS DR2, \citealt{Chambers et al.(2016)}). Any catalogs associated rejection table was not used during the query for photometric data. The associated photometric bands used in our analysis were W1 and W2 from CatWISE2020, W3 and W4 from $AllWISE$, J and $K_{s}$ from VHS (J and $K_{MKO}$ from UKIDSS LAS DR9/UHS, and $g_{ps}$, $r_{ps}$, $i_{ps}$, $z_{ps}$, $y_{ps}$ from PanSTARRS DR2. All photometric data are provided in Table \ref{Table 2}. 

\begin{deluxetable*}{clc}[ht]
\tabletypesize{\scriptsize}
\tablecaption{Discoverer, Photometry, Astrometry, and Derived Quantities}\label{Table 2}  
\tablehead{
\colhead{Column Name}     & \colhead{Column Description}  & \colhead{Example Entry} \\
}
\startdata
CWISE\_Name         & CWISE Designation                                                                                    & J043918.82+134231.1 \\
CW\_RA                  & CWISE Right Ascension measurement (deg)                           & 1.05837 \\
CW\_DEC                 & CWISE Declination measurement (deg)                                  & 4.73992\\
CW\_RAerr                 & CWISE Right Ascension uncertainty measurement (deg)                                  & 0.07510\\
CW\_DECerr                 & CWISE Declination Uncertainty measurement (deg)                                  & 0.07380\\
CW\_MJD                 & CWISE Modified Julian Date                                   & 56700\\
VHS\_RA                  & VHS Right Ascension measurement (deg)                           & ... \\
VHS\_DEC                 & VHS Declination measurement (deg)                                  & ...\\
VHS\_MJD                 & VHS Modified Julian Date                                  & ...\\
LAS\_RA                  & UKIDSS LAS DR9 Right Ascension measurement (deg)                           & 1.05860 \\
LAS\_DEC                 & UKIDSS LAS DR9 Declination measurement (deg)                                  & 4.74035\\
LAS\_MJD                 & UKIDSS Modified Julian Date                                  & 55187\\
UHS\_RA                  & UHS Right Ascension measurement (deg)                           & ... \\
UHS\_DEC                 & UHS Declination measurement (deg)                                  & ...\\
UHS\_MJD                 & UHS Modified Julian Date                                  & ...\\
PS\_RA                  & PanSTARRS DR2 Right Ascension measurement (deg)                           & 1.05841 \\
PS\_DEC                 &  PanSTARRS DR2 Declination measurement (deg)                                  & 4.73997 \\
PS\_RAerr                 &  PanSTARRS DR2 Right Ascension uncertainty measurement (deg)                                  & 0.12252\\
PS\_DECerr                 &  PanSTARRS DR2 Declination uncertainty measurement (deg)                                  & 0.12252\\
PS\_MJD                 &  PanSTARRS DR2 Modified Julian Date                                        & 56873\\
Discover            & Object Discoverers    & Yadukrishna Raghu, ...  \\
W1                  & W1-band magnitude from CatWISE2020 (mag)                                                             & 14.873              \\
W1err               & Uncertainty in W1, as provided by CatWISE2020 (mag)                                                  & 0.017               \\
W2                  & W2-band magnitude from CatWISE2020 (mag)                                                             & 14.647              \\
W2err               & Uncertainty in W2, as provided by CatWISE2020 (mag)                                                  & 0.024               \\
W3                  & W3-band magnitude from \textit{AllWISE} (mag)                                       & 11.777              \\
W3err               & Uncertainty in W3, as provided by \textit{AllWISE} (mag)                            & …                   \\
W4                  & W4-band magnitude from \textit{AllWISE} (mag)                                       & 8.621               \\
W4err               & Uncertainty in W4, as provided by \textit{AllWISE} (mag)                            & …                   \\
J                   & J-band magnitude from VHS/UKIDSS LAS DR9 (mag)                                                       & 16.800           \\
Jerr                & Uncertainty in J, as provided by VHS/UKIDSS LAS DR9 (mag)                                            & 0.019            \\
Kmko                & K$_{MKO}$-band magnitude from UKIDSS LAS DR9 (mag)                                                   & 15.609           \\
Kmkoerr             & Uncertainty in K$_{MKO}$, as provided by UKIDSS LAS DR9 (mag)                                        & 0.022            \\
Ks                  & K$_{s}$-band magnitude from VHS (mag)                                                                & …                   \\
Kserr               & Uncertainty in K$_{s}$, as provided by VHS (mag)                                                     & …                   \\
g                   & g-band magnitude from PanSTARRS DR2 (mag)                                                            & …                   \\
gerr                & Uncertainty in g, as provided by PanSTARRS DR2 (mag)                                                 & …                   \\
r                   & r-band magnitude from PanSTARRS DR2 (mag)                                                            & …                   \\
rerr                & Uncertainty in r, as provided by PanSTARRS DR2 (mag)                                                 & …                   \\
i                   & i-band magnitude from PanSTARRS DR2 (mag)                                                            & …                   \\
ierr                & Uncertainty in i, as provided by PanSTARRS DR2 (mag)                                                 & …                   \\
z                   & z-band magnitude from PanSTARRS DR2 (mag)                                                            & 20.069         \\
zerr                & Uncertainty in z, as provided by PanSTARRS DR2 (mag)                                                 & 0.028         \\
y                   & y-band magnitude from PanSTARRS DR2 (mag)                                                            & 18.867         \\
yerr                & Uncertainty in y, as provided by PanSTARRS DR2 (mag)                                                 & 0.034         \\
CW\_PMRA                & Proper motion in R.A., as provided by CatWISE2020 (arcsec yr$^{-1}$)                               & 0.321             \\
CW\_PMRAerr             & Uncertainty in PMRA, as provided by CatWISE2020 (arcsec yr$^{-1}$)                                 & 0.013              \\
CW\_PMDec               & Proper motion in decl., as provided by CatWISE2020 (arcsec yr$^{-1}$)                              & 0.134               \\
CW\_PMDecerr            & Uncertainty in PMDec, as provided by CatWISE2020 (arcsec yr$^{-1}$)                                & 0.013              \\
CW\_PMTot               & Total proper motion, as provided by CatWISE2020 (arcsec yr$^{-1}$)                                 & 0.348              \\
CW\_PMToterr            & Uncertainty in total proper motion, as provided by CatWISE2020 (arcsec yr$^{-1}$)                  & 0.018              \\
PMRA                & Proper motion in R.A., as described in \S\ref{photometry} (arcsec yr$^{-1}$)                               & -0.166             \\
PMRAerr             & Uncertainty in PMRA, as described in \S\ref{photometry} (arcsec yr$^{-1}$)                                 & 0.039              \\
PMDec               & Proper motion in decl., as described in \S\ref{photometry} (arcsec yr$^{-1}$)                              & -0.321               \\
PMDecerr            & Uncertainty in PMDec, as described in \S\ref{photometry} (arcsec yr$^{-1}$)                                & 0.061              \\
PMTot               & Total proper motion, as described in \S\ref{photometry} (arcsec yr$^{-1}$)                                 & 0.362              \\
PMToterr            & Uncertainty in total proper motion, as described in \S\ref{photometry} (arcsec yr$^{-1}$)                  & 0.057              \\
SpT                 & Derived estimated photometric spectral type from \S\ref{chi2}                                        & L3                  \\
SpT\_spec           & Derived spectral type from observed spectrum in \S\ref{spectrum}                                     & ...                  \\
chi2                & Derived $\chi^2$ value from \S\ref{chi2}                             & 5.660               \\
distance            & Derived distance from \S\ref{distance} (pc)                          & 64.820              \\
distance\_err       & Uncertainty in distance, as derived from \S\ref{distance} (pc)       & 2.890               \\
vtan                & Derived V$_{tan}$ from \S\ref{distance} (km s$^{-1}$)                    & 106.809             \\
vtan\_err           & Uncertainty in vtan, as derived from \S\ref{distance} (km s$^{-1}$)      & 7.356               \\
\enddata
\tablecomments{This table only describes the columns in the full table. The full table is available from the journal.}
\end{deluxetable*}

To gather these data and their associated uncertainties, we used WRAP with a search radius of 150$\arcsec$, this large search radius is to aid visual confirmation in crowded fields. We confirmed the photometric data visually using WRAP's catalog overlay feature, which overlays catalog data on the relevant catalog imaging. WRAP also provides a WiseView (\citealt{Caselden et al.(2018)}) popup that helped locate each \texttt{SMDET} discovery in the relevant catalog data. 

Using WRAP, we collected Right Ascension (R.A.), Declination (Decl.), and Modified Julian Date (MJD) measurements from CatWISE2020, VHS, UKIDSS LAS DR9, UHS, and PanSTARRS DR2 for each object. We then fit a linear regression line to these measurements to determine the proper motions and associated uncertainties for each object. Due to the lack of R.A. and Decl. uncertainties in VHS, UKIDSS LAS DR9, and UHS, we required a minimum of three data points per object to obtain proper motions and their uncertainties. For objects with insufficient data points, we relied on CatWISE2020 astrometric data, as our candidates were initially discovered using \textit{WISE} W1 and W2 imaging. All astrometric data is provided in Table \ref{Table 2}.

One object, CWISE J024822.37+674812.6 (hereafter, J0248+6748), stood out as interesting but lacked $J$-band photometry, leading us to gather $J$-band photometry using the \textit{Wide-field InfraRed Camera} (hereafter, WIRC; \citealt{Wilson et al.(2003)}) on the Palomar 200” telescope. The target was observed on 26 July 2022 (UT), with clear skies, no clouds, and a seeing full-width at half-maximum (FWHM) of 4 pixels (1\arcsec). The total integration time was 30 minutes, divided into 2-minute exposures obtained with a 15-point dithering pattern. Each 2-minute exposure is further divided into 4 frames of 30s each. WIRC has a pixel scale of 0\farcs2487 per pixel and a total field of view of 8\farcm7 $\times$ 8\farcm7. The data were reduced using custom scripts written in the Interactive Data Language (IDL\footnote{\url{https://www.nv5geospatialsoftware.com/Products/IDL}.}) that included dark subtraction, flat fielding, sky subtraction, and image stacking for the final mosaic. The mosaic was astrometrically and photometrically calibrated using stars from 2MASS (\citealt{Skrutskie et al.(2006)}). Aperture photometry was performed on J0248+6748 using an aperture of 6.5 pixels (1\farcs6), $\sim$1.5 times the seeing FWHM.

Figure \ref{Figure 2} is a compilation of color-color and color-reduced proper motion diagrams, with PanSTARRS 3$\pi$ survey objects (\citealt{Best et al.(2018)}) shown as colored data points in the background. Reduced proper motion at W2 ($H_{W2}$) is a stand-in for absolute brightness in the absence of measured trigonometric parallaxes, and is defined as
\begin{equation}
    H_{W2} = W2 + 5\times \log_{10}(\mu_{total}) + 5 =  M_{W2} + 5\times \log_{10}(V_{tan}) + 1.62
\end{equation}, where W2 and $M_{W2}$ are  measured in magnitudes, $\mu_{total}$ is measured in arcsec yr$^{-1}$, and $V_{tan}$ is measured in km s$^{-1}$. We see in Figures \ref{Figure 2} that all but three objects follow the general trend from \cite{Best et al.(2018)}. The three outliers, marked with numbers, are later discussed in \S\ref{cheetoo}. 

\section{Spectral Type Estimates} \label{spectral}

\subsection{Photo-Type Spectral Type Estimation Method}\label{chi2}

Spectral types can be estimated from photometry using the photo-type spectral type estimation technique defined in \cite{Skrzypek et al.(2015),Skrzypek et al.(2016)}. The method calculates the inverse variance-weighted difference between observed and empirical template colors as a function of spectral subtype:
\begin{equation}\label{eq1}
\hat{m}_{B,t} = \frac{\sum_{b=1}^{N_b} \frac{\hat{m}_{b} - c_{b,t}}{\sigma_{b}^2}}{\sum_{b=1}^{N_b} \frac{1}{\sigma_{b}^2}}
\end{equation}
Here, $\{ t \}$ is an array of spectral types, ranging over M0-T8 in full subtype increments; $\{ b \}$ is an array of $N_{b}$ photometric bands as listed in \S\ref{photometry}; $B$ is the reference band, chosen as $J_{MKO}$; $\hat{m}_{b}$ $\pm$ $\sigma_{b}$ are the measured magnitudes and uncertainties; $c_{b,t}$ are the empirical spectral subtype template color, defined as $m_b-J_{MKO}$ and drawn from \cite{Skrzypek et al.(2015)}; and $\hat{m}_{B,t}$ is the averaged synthetic $J_{MKO}$ apparent magnitude at each spectral subtype.

\begin{figure}[ht]
        \centering
        \includegraphics[scale = 0.28]{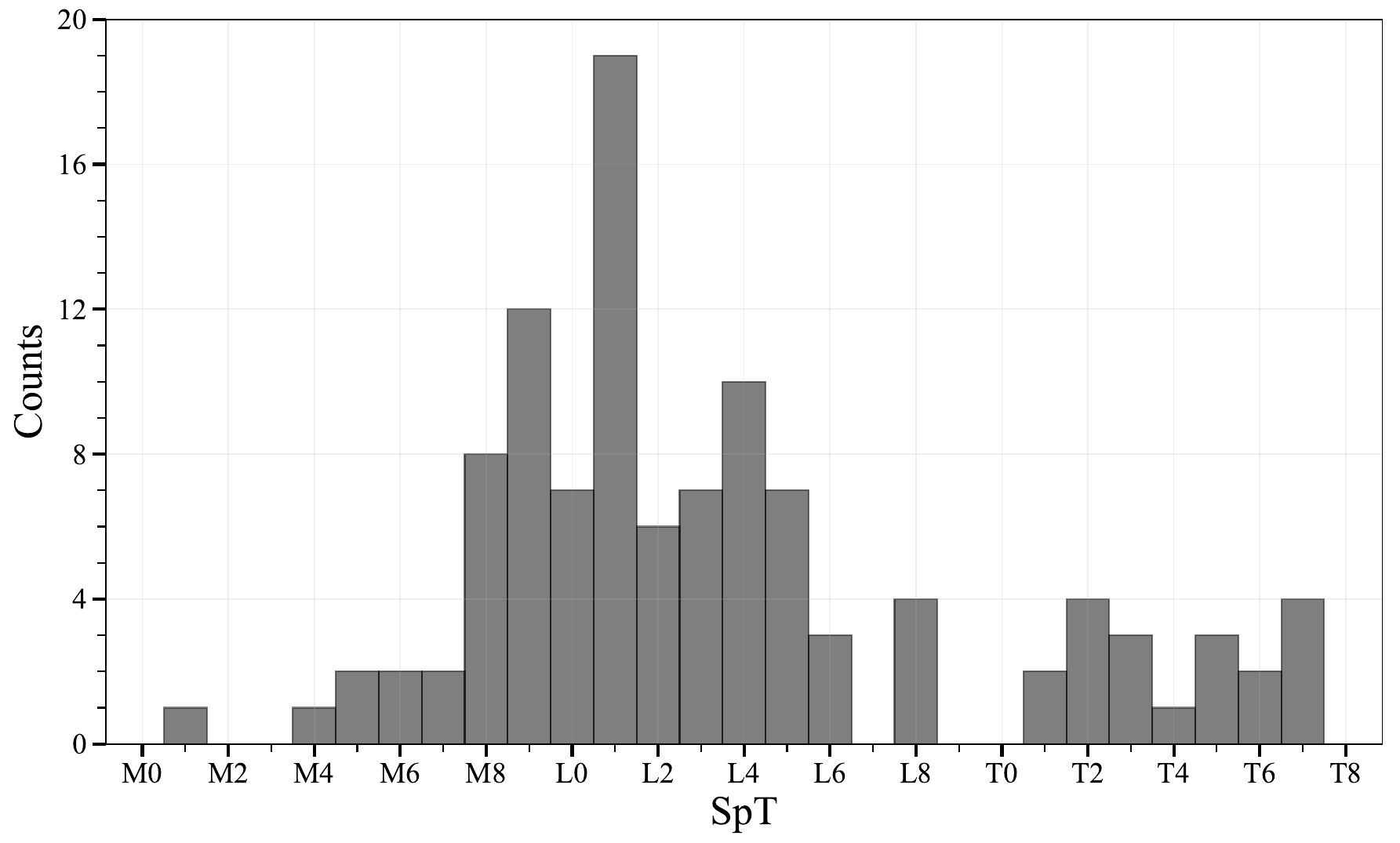}
        \caption{Distribution of spectral type estimates for our \texttt{SMDET} sample. This figure does not include subdwarf candidates discovered in this paper. }
        \label{Figure 9}
\end{figure}

\begin{figure}[ht]
        \centering
        \includegraphics[scale = 0.2]{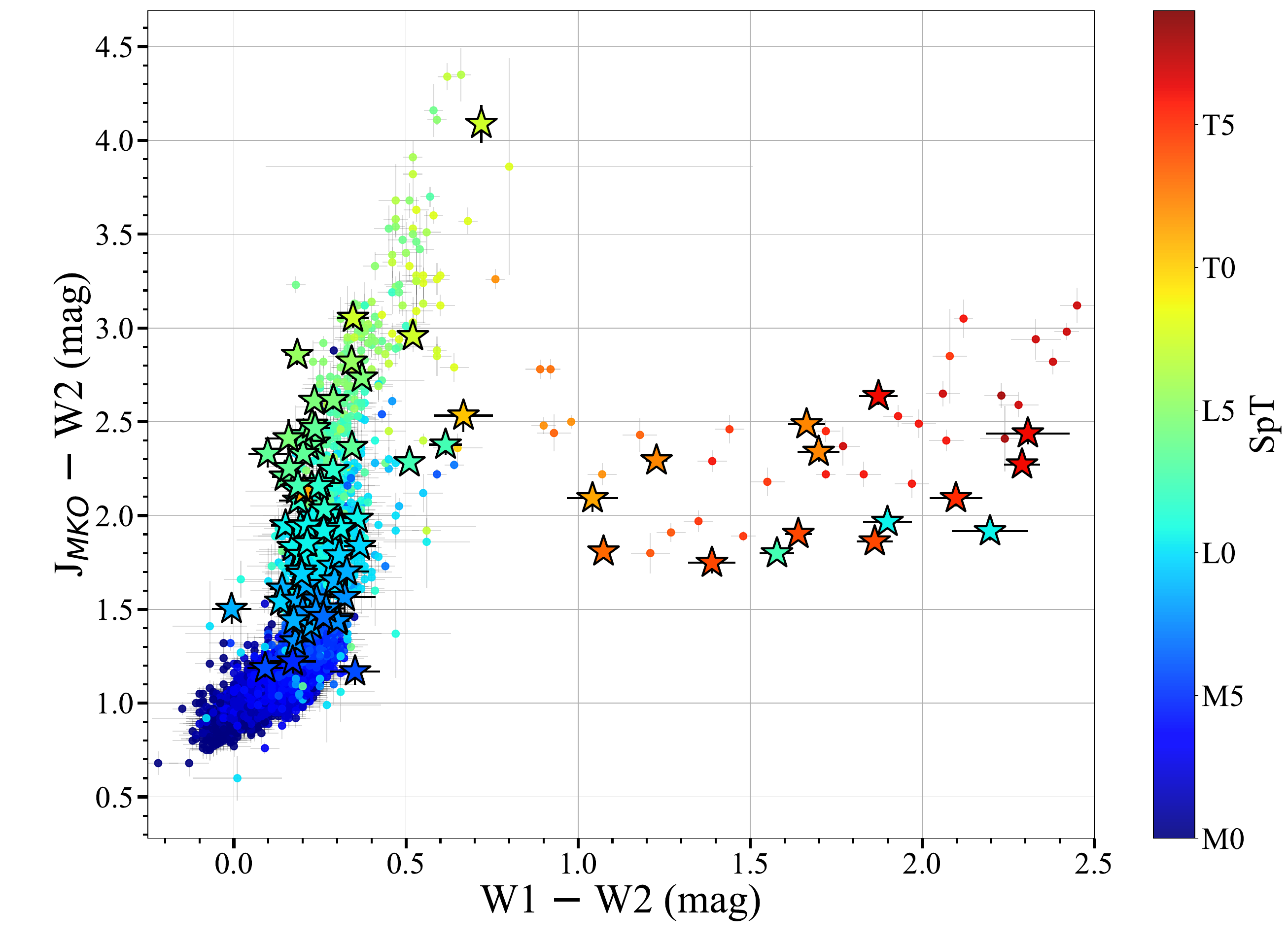}
        \caption{A color-color diagram illustrating our photo-type spectral type estimates compared to known spectral types from \cite{Best et al.(2018)}. The starred objects are the 89 with available J$_{MKO}$ out of 110 photo-type spectral estimates from this study. }
        \label{Figure 3}
\end{figure}

\begin{figure*}[ht]
        \centering
        \includegraphics[scale = 0.420]{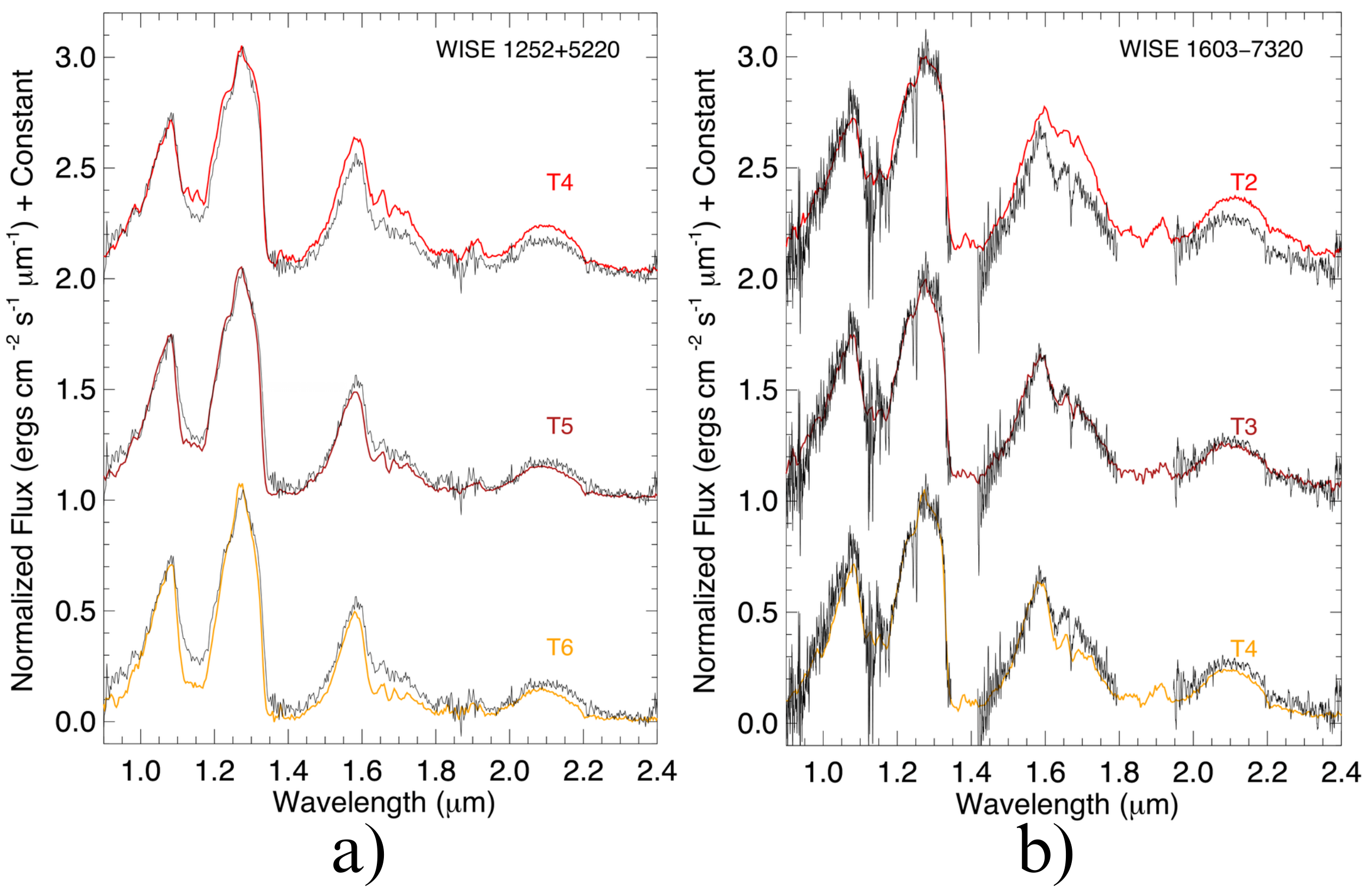}
        \caption{ a) The near-infared (0.8--2.4~$\mu$m) SpeX spectrum of J1252+5220 (black lines) compared to T4, T5, and T6 spectral standards (orange, maroon, and yellow lines).
        b) Similar sequence comparing ARCoIRIS data for J1603$-$7320 to T2, T3, and T4 spectral standards. The standard spectra are those defined in \citet{2006ApJ...637.1067B}, with data from \citet{2004AJ....127.2856B} and \citet{2006ApJ...639.1095B}. All are normalized at the peak of the $J$ bandpass ($\sim$1.3 microns).}
        \label{Figure 4}
\end{figure*}

The best spectral type match was inferred using a $\chi^2$ statistic,
\begin{equation}\label{eq2}
\chi^2({\{ \hat{m}_{b} \}},{\{ \sigma_{b} \}}, \hat{m}_{B,t}, t) = \sum_{b=1}^{N_b} (\frac{\hat{m}_b - \hat{m}_{B,t} - c_{b,t}}{\sigma_{b}^2})^2
\end{equation}
where lower $\chi^2$ indicates better matches. This methodology is further detailed in \cite{Skrzypek et al.(2015)} Appendix A. \cite{Tang et al.(2018)} further developed this method into a Python package, \texttt{phot-d}, that was provided to the authors (Tang, priv.\ comm., 2022). The package use spectral type-magnitude templates from \cite{Pecaut Mamajek(2013)} for B2 to M4, \cite{Best et al.(2018)} for M6 to L9, and \cite{Skrzypek et al.(2015)} for T0 to T8. Note that M5 is missing in the template set. Due to the absence of Y dwarfs in these datasets, we regard any object exhibiting a color of W1$-$W2 \textgreater 3.0 mag as a Y dwarf candidate (Figure 17h of \citealt{Kirkpatrick et al.(2021)}), although no such candidates were identified in our search. Note that spectral type uncertainties were not calculated, thus they were assumed to be ideal.

Of the 118 candidates in our sample, 110 have spectral type estimates from this technique, and are listed in Table \ref{Table 2} and their distribution displayed in Figure \ref{Figure 9}. Out of the 110 photometric spectral type estimates, 6 objects were found to have spectral types earlier than M7, indicating they are not ultracool dwarfs (M7 and later). Additionally, the five objects without spectral type estimates have a W1$-$W2 (mag) color between 0.55 and 0.71, which suggests they may be late-L dwarfs (\citealt{Kirkpatrick et al.(2021)}). Lastly, 3 objects were discovered to be candidate subdwarfs and are later discussed in \S\ref{cheetoo}. Spectroscopic observations are necessary for all objects in this study to confirm their estimated spectral types.

The validity of our photometric classifications is illustrated in Figure \ref{Figure 3}, which maps the spectral types of both the \texttt{SMDET} sample and the \cite{Best et al.(2018)} sample on a W1$-$W2 (mag) vs. $J-$W2 (mag) diagram. The photometric classifications align well with spectral types for most of the sample, but for some late-type T dwarfs the photometric classifications are up to a full class too early. Further spectral analysis is needed to determine why these sources show such large deviations between classification methods.

\subsection{Observed Candidate Spectra}\label{spectrum}

We acquired spectra for two objects, CWISE J125247.68+522015.5 (hereafter, J1252+5220) and CWISE J160347.72$-$732054.3 (hereafter, J1603$-$7320), to validate our spectral type estimates. These sources have photometric classifications of T5 and T4, respectively.

We observed J1252+5220 on the night of 22 January 2019 (UT) under good conditions with clear skies using the SpeX spectrograph \cite{2003PASP..115..362R} on NASA’s IRTF telescope. These data were taken in prism mode using the 0.8$\arcsec$ slit to achieve a resolving power of $\sim$75 over the 0.8-2.5$\mu$m wavelength range. We obtained 6 AB nods using 180s exposures on the target and then acquired the A0 star HD 99966 for telluric correction using 0.1s exposures and 10 AB nods. All data were reduced using the Spextool package (\citealt{Cushing et al.(2004)}) with telluric correction and flux calibration of the A0 star following the technique described in \cite{Vacca et al.(2003)}. 

We observed J1603$-$7320 on the night of 01 April 2018 (UT) under good conditions with clear skies from the Astronomy Research using the Cornell Infrared Imaging Spectrograph (ARCoIRIS; \citealt{2004SPIE.5492.1295W}) on the 4m Blanco telescope located at the Cerro Tololo Inter-American Observatory (CTIO). ARCoIRIS takes simultaneous spectra across six cross-dispersed orders covering the 0.8–2.4$\mu$m range, with a resolving power of $\sim$3,500 using the 1.0$\arcsec$ slit. We obtained two different nod positions across the slit and acquired 8 AB exposures of 180s each. After taking science exposures, we obtained nine calibration lamp images followed by an A0 star, HD 149954, for telluric correction by taking 8 AB exposure nodes of 10s each. All data were reduced similarly to above. 

We visually fit the spectral templates from \citet{2006ApJ...637.1067B}, with data from \citet{2004AJ....127.2856B} and \citet{2006ApJ...639.1095B}, to our measured spectra. These data provide spectroscopic classifications of T5 for J1252+5220 and T3 for J1603$-$7320, matching the photometric estimates within one subtype (Figure \ref{Figure 4}). Although a larger sample of spectra needs to be gathered to verify classifications over a wider range of types, these two examples provide a promising example of validation.

\section{Distance and Tangential Velocity Estimates}\label{distance}
\begin{figure}[ht]
        \centering
        \includegraphics[scale = 0.2]{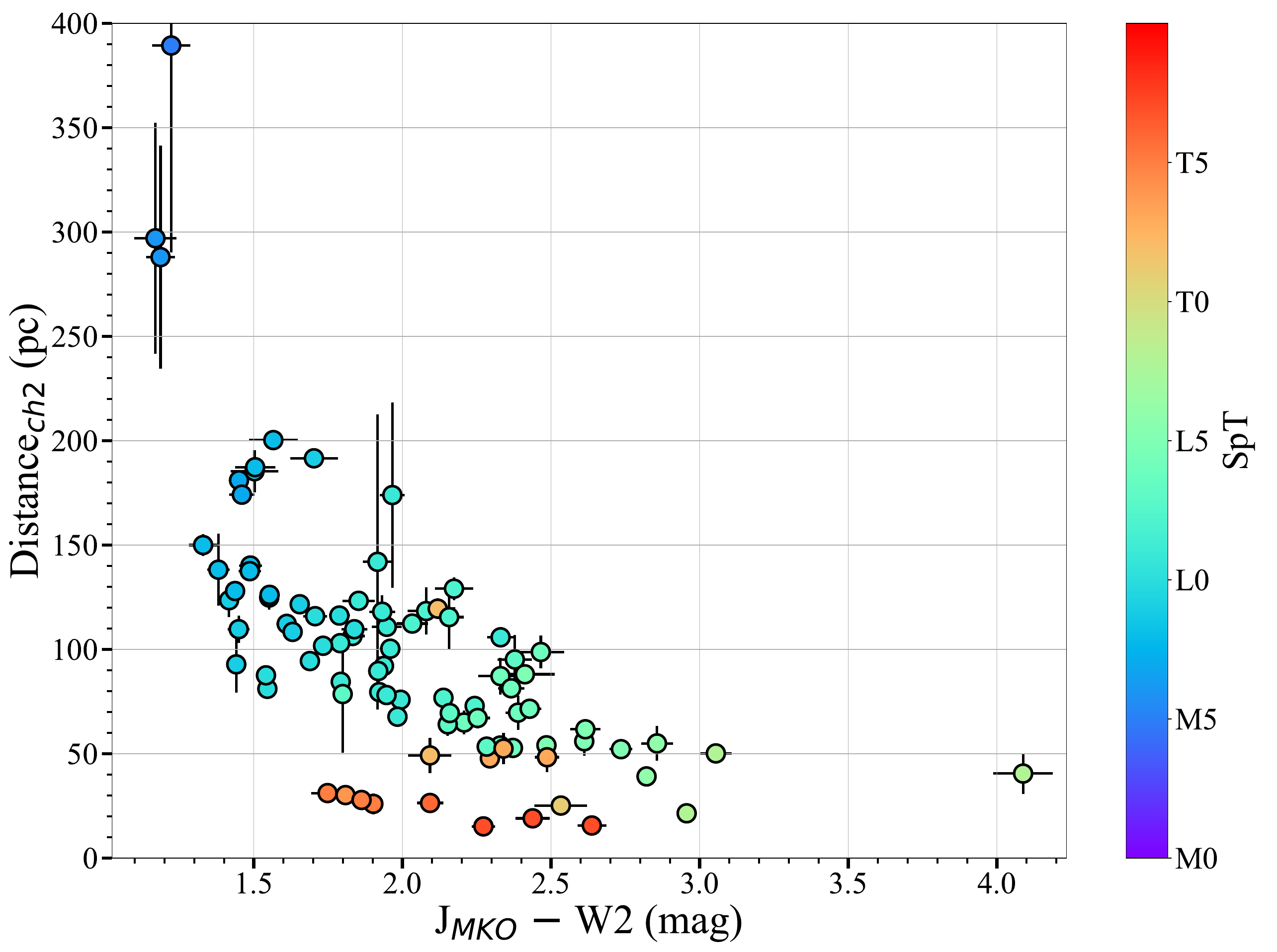}
        \caption{$J_{MKO}-$W2 versus estimated distance with associated uncertainties. Symbol color encodes photometric classifications.}
        \label{Figure 5}
\end{figure}

The photo-type code uses absolute magnitudes for its empirical templates, and these can be used to estimate distances for our sources and their associated uncertainties. We used the M$_{W2}$ empirical absolute magnitudes and $W2$ measured magnitudes to calculate distances, ignoring reddening effects, which are provided in Table \ref{Table 2}. Using a standard error propagation equation, the associated uncertainties were derived from absolute magnitude template errors combined with apparent magnitude errors. Figure \ref{Figure 5} displays $J_{MKO}-$W2 (mag) versus estimated distance (pc), demonstrating the expected trend wherein objects with later spectral types are only seen if they are closer to the Sun due to the photometric detection limit of \textit{WISE}. 

\begin{figure}[ht]
        \centering
        \includegraphics[scale = 0.2]{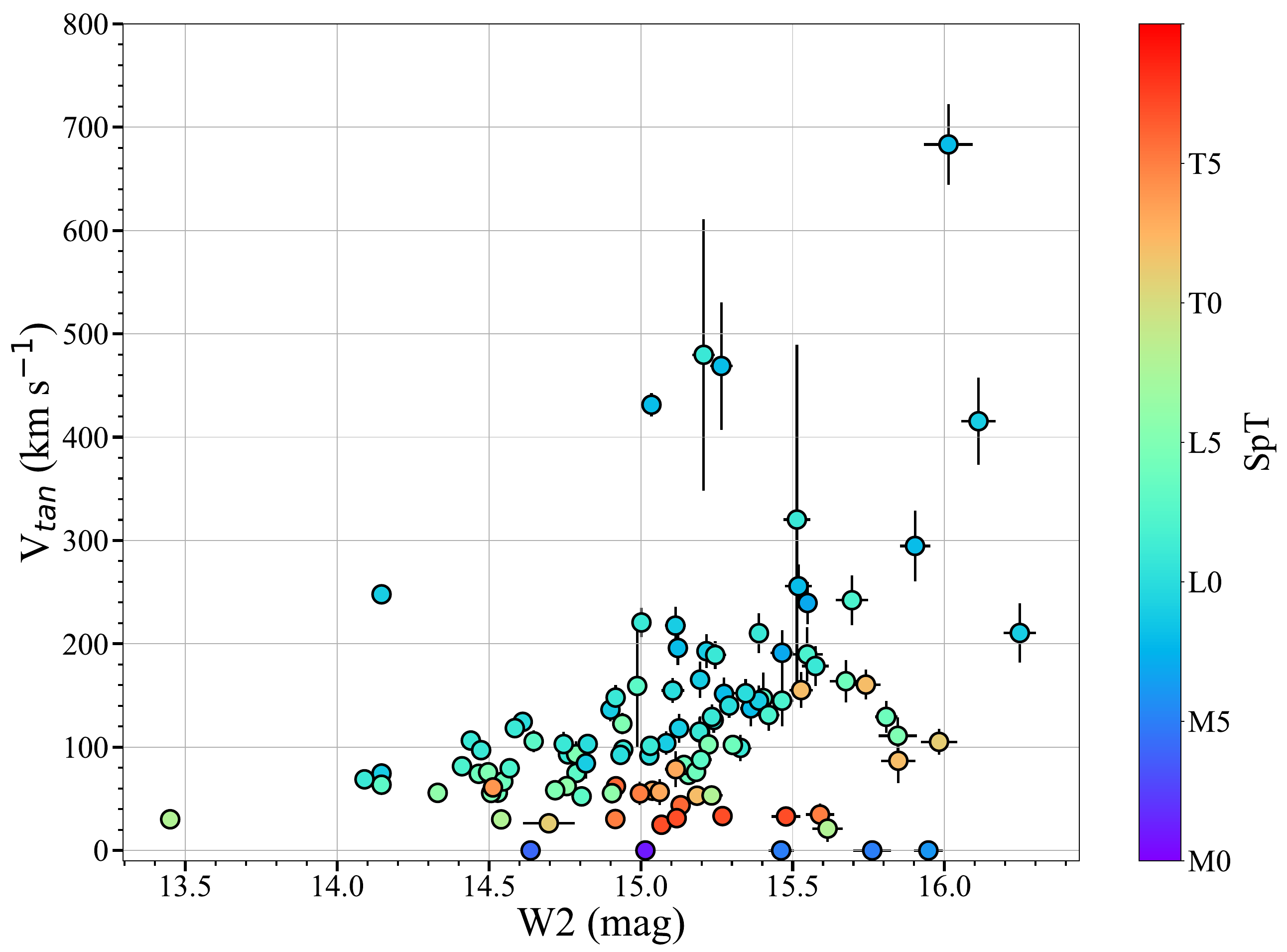}
        \caption{Apparent W2 versus estimated tangential velocity with associated uncertainties. Symbol color encodes photometric classifications.}
        \label{Figure 6}
\end{figure}

With the estimated distances and measured proper motions (where calculated proper motions were used when possible) we calculated tangential velocity ($V_{tan}$) estimates, also provided in Table \ref{Table 2}. Figure \ref{Figure 6} compares apparent W2 (mag) versus $V_{tan}$ (km s$^{-1}$). The tangential velocity uncertainties were calculated using standard error propagation, errors from the total proper motion and distance. Some of the tangential velocities are extremely high, exceeding the escape velocity of the Milky Way (\citealt{Koppelman Helmi(2021)}), albeit with large uncertainties. Measured parallaxes are needed to verify these extreme velocities. 

\begin{deluxetable*}{cllllccc}[ht]
\tabletypesize{\scriptsize}
\tablecaption{Subdwarf Candidates}\label{Table 3}  
\tablehead{
\colhead{Figure \ref{Figure 2} $\#$}     & \colhead{Object Name}  & \colhead{ R.A. (deg)}    & \colhead{Decl. (deg)} & \colhead{SpT} & \colhead{$M_J$ (mag)} & \colhead{Distance (pc)} & \colhead{$V_{tan}$ (km s$^{-1}$}) \\
}
\startdata
1 & CWISE J024822.37+674812.6 & 42.0932152 & 67.8035141 & esdT?  & $\sim$13   & $\sim$47 & $\sim$223 \\
2 & CWISE J052544.47+443409.9 & 81.4353002 & 44.5694241 & sdT8?  & $\sim$13.5 & $\sim$27 & $\sim$66  \\
3 & CWISE J120642.54+762641.6 & 181.677282 & 76.4449143 & esdT?  & $\sim$13   & $\sim$58 & $\sim$293 \\
\enddata
\end{deluxetable*}

Objects CWISE J194427.98+053342.6 and CWISE J203710.21+545750.8 are excluded from Figures \ref{Figure 5} and \ref{Figure 6} due to their exceptionally large distances and tangential velocities estimates of 509 (pc)/2223 (km s$^{-1}$) and 808 (pc)/1299 (km s$^{-1}$), respectively, which exceed the scale intended for these figures. These extraordinarily high values should be interpreted with caution and verified through trigonometric parallax measurements.

Six objects, all of which are earlier than M7 and thus not ultracool dwarfs, either lacked reliable distance estimates or had substantial errors. These objects are J030138.04$-$401755.2, J153103.62+130935.2, J194427.98+053342.6, J203710.21+545750.8, J221039.79$-$313533.2, and J222437.97$-$163134.6. The discrepancies are likely due to poor fits between the template absolute magnitudes and the measured data, coming from the minimal color differences among early spectral types. To address this issue, we recalculated the distance and tangential velocity estimates for these objects using the spectral type versus absolute W2 magnitudes from Table 7 of \cite{Cifuentes et al.(2020)}. This adjustment significantly improved the estimates. Nevertheless, measured trigometric parallaxes are still necessary to verify these estimates.

\section{Candidate Subdwarfs}\label{cheetoo}
With any kinematic selection, there is a higher probability of finding subdwarfs (\citealt{Arifyanto et al.(2005)}). Here, we assess the presence of T subdwarfs (sdT) and extreme T subdwarfs (esdT; \citealt{Schneider et al. 2020}) in our sample given our focus on later-type brown dwarfs. We based our analysis on a compilation of 9 color-color diagrams shown in Figure 8 of \cite{Zhang et al.(2019)}, and the color-color box defined in Figure 1 of \cite{Meisner et al.(2021)}, to identify sdT and esdT candidates, respectively. The latter region encompasses 1.1 \textless W1$-$W2 (mag) \textless 1.75 and $J-$W2 (mag) \textgreater 3. We identified three possible subdwarfs, listed in Table \ref{Table 3}. 

CWISE J052544.47+443409.9 (hereafter, J0525+4434) exhibits colors akin to those of a $\sim$sdT8? candidate reported in Figure 8f and 8i of \cite{Zhang et al.(2019)}, which we tentatively assign as the classification of this source. J0525+4434 has a substantial proper motion of 0{\farcs}52$\pm$0{\farcs}13 arcsec yr$^{-1}$. Using the $M_{W2}$ reported in \cite{Zhang et al.(2019)}, $M_{W2}$$\sim$13.5 mag, we estimate a distance of 27 pc for this source, corresponding to $V_{tan}$ $\approx$ 66 km s$^{-1}$.

Two sources, CWISE J024822.37+674812.6 (hereafter, J0248+6748) and CWISE J120642.54+762641.6 (hereafter, J1206+7626) fall within the extreme T subdwarf region defined in \cite{Meisner et al.(2021)}. In the absence of spectral data, we adopt preliminary classifications of $\sim$esdT? for both objects. Both exhibit very large proper motions of 1{\farcs}00$\pm$0{\farcs}13 yr$^{-1}$ and 1{\farcs}07$\pm$0{\farcs}14 yr$^{-1}$, respectively. \cite{Brooks et al.(2022)} adopt an absolute magnitude of $M_{W2}$$\sim$13 mag for esdT? candidates within a similar region of color-color space, implying distance estimates of $\sim$47 pc for J0248+6748 and $\sim$58 pc for J1206+7626, which in turn imply $V_{tan}$ $\sim$223 km s$^{-1}$ and $\sim$293 km s$^{-1}$, respectively. Such high velocities are consistent with halo population objects; however, spectra and parallax measurements are needed to validate these measures.

\section{Candidate Young Objects}\label{young}
Among the few sources in our sample which did not have reliable photometric classifications, but are not subdwarfs, it is possible that their deviant colors arise from having low surface gravity atmospheres. We used the BANYAN $\Sigma$ Bayesian classifier (\citealt{Gagne et al.(2018b)}) to examine whether any of our discoveries are candidate members of a known nearby young association. The BANYAN $\Sigma$ tool uses available observational properties such as sky position, proper motion, radial velocity, and distance to assess whether a given object has $XYZ$ Galactic space coordinates and $UVW$ space velocities that match the location and kinematics of known young associations. In cases where a radial velocity and/or a distance is missing, these parameters can be marginalized and a membership probability can still be determined.
 
We used the photometric distances presented in Table \ref{Table 2} combined with the sky coordinates and proper motions from Table \ref{Table 2} to determine membership probabilities. We used the updated kinematic models of \cite{Gagne(2024)}, up to a distance of 814 pc corresponding to the largest plausible distance of our ultracool dwarf candiates at 3$\sigma$. We used a Monte Carlo approach with 10$^4$ samples to marginalize over the measurement errors of the proper motion, which is computationally slower but more accurate than the error propagation of proper motion otherwise done in BANYAN $\Sigma$ (\citealt{Gagne et al.(2018b)}). Using this method, we identified only two low-likelihood candidate members among our sample both of which lack a photometric classification. 
 
CWISE J060938.14$-$542916.3 (hereafter J0609$-$5429) was found to be a low-probability (50$\%$) candidate member of a nearby moving group. This probability is comprised of: 59$\%$ (equivalent to 29.5$\%$ overall) of the membership probability is assigned to the $\approx$300 Myr-old corona of Group 23 from \cite{Oh et al.(2017)} (see \citealt{Moranta et al.(2022)} for a discussion of this structure, and \citealt{Kounkel et al.(2020)} for the age estimation of the associated structure Theia 599), and 36$\%$ (18$\%$ overall) is associated with the $\approx$700 Myr-old (\citealt{Galindo-Guil et al.(2022)}) Hyades tidal tail (\citealt{Roser et al.(2019)}). The remaining 5$\%$ (2.5$\%$ overall) probability is assigned to the field.
 
CWISE J141615.86+685919.7 (hereafter J1416+6859) was found to be a high-probability (85$\%$) candidate member of a nearby moving group. This is split between two groups: 68$\%$ (equivalent to 57.8$\%$ overall) in the $\approx$500-600 Myr-old Oceanus moving group (\citealt{Gagne et al.(2023)}), and 24$\%$ (20.4$\%$ overall) in the $\approx$133 Myr-old AB Doradus moving group (\citealt{Zuckerman et al.(2004)}, \citealt{Gagne et al.(2018b)}). The remaining 8$\%$ (6.8$\%$ overall) probability is assigned to the field.
 
We consider both of these objects to be weak candidate members of their respective associations due to their lack of classifications, photometric distance estimates, and tangential velocity estimates, and therefore complete assessment of their $XYZ$ and $UVW$ kinematics. This means that the probabilities are based solely on the sky position and proper motion of these candidates. We also lack spectral evidence of youth which would be required to make them convincing members of their respective associations. Additionally, our reliance on the photometric distances for other sources in our sample could have caused us to miss members of nearby young associations if they are young enough to have atypical colors or absolute magnitudes (usually $\leq$ 200 Myr, \citealt{Faherty et al.(2016)}). Therefore, once trigonometric parallaxes are taken for these objects, the BANYAN $\Sigma$ analysis should be performed again. 

Alongside the BANYAN $\Sigma$ calculations, we found that CWISE J153653.38+253801.9 (herafter J1536+2538) has J$-$K and J$-$W2 colors that are extremely red for its estimated spectral type. This may be a result of youth, as young L dwarfs are known to possess redder near-infrared colors than their field-age counterparts as they are still contracting and therefore have low surface gravities. The resulting low atmospheric pressure causes reduced collision-induced absorption by H$_2$ \citep{Linsky1969} and an excess of high-altitude dust clouds in their photospheres \citep{Cushing2008,Faherty et al.(2016)}, shifting emergent flux to longer wavelengths. However, BANYAN $\Sigma$ indicates that J1536+2538 is not a member of a known young moving group. There is a possibility that this source is a young object that has been ejected from its natal environment. However, a population of field-age objects exists that possess red near-infrared colors (\citealt{Marocco2014}), thought to have an excess of sub-micron sized dust grains in their upper atmospheres \citep{Hiranaka2016}. Other factors are also known to redden the near-infrared colors of brown dwarfs, such as supersolar metallicity \citep{Looper2008} and equator-on inclination angle \citep{Vos2017,Suarez2022,Suarez2023}. Consequently, spectroscopic data must be obtained for J1536+2538 for confirmation of its youth.

\section{Limitations of This Study}\label{dis}

\begin{figure}[ht]
        \centering
        \includegraphics[scale = 0.18]{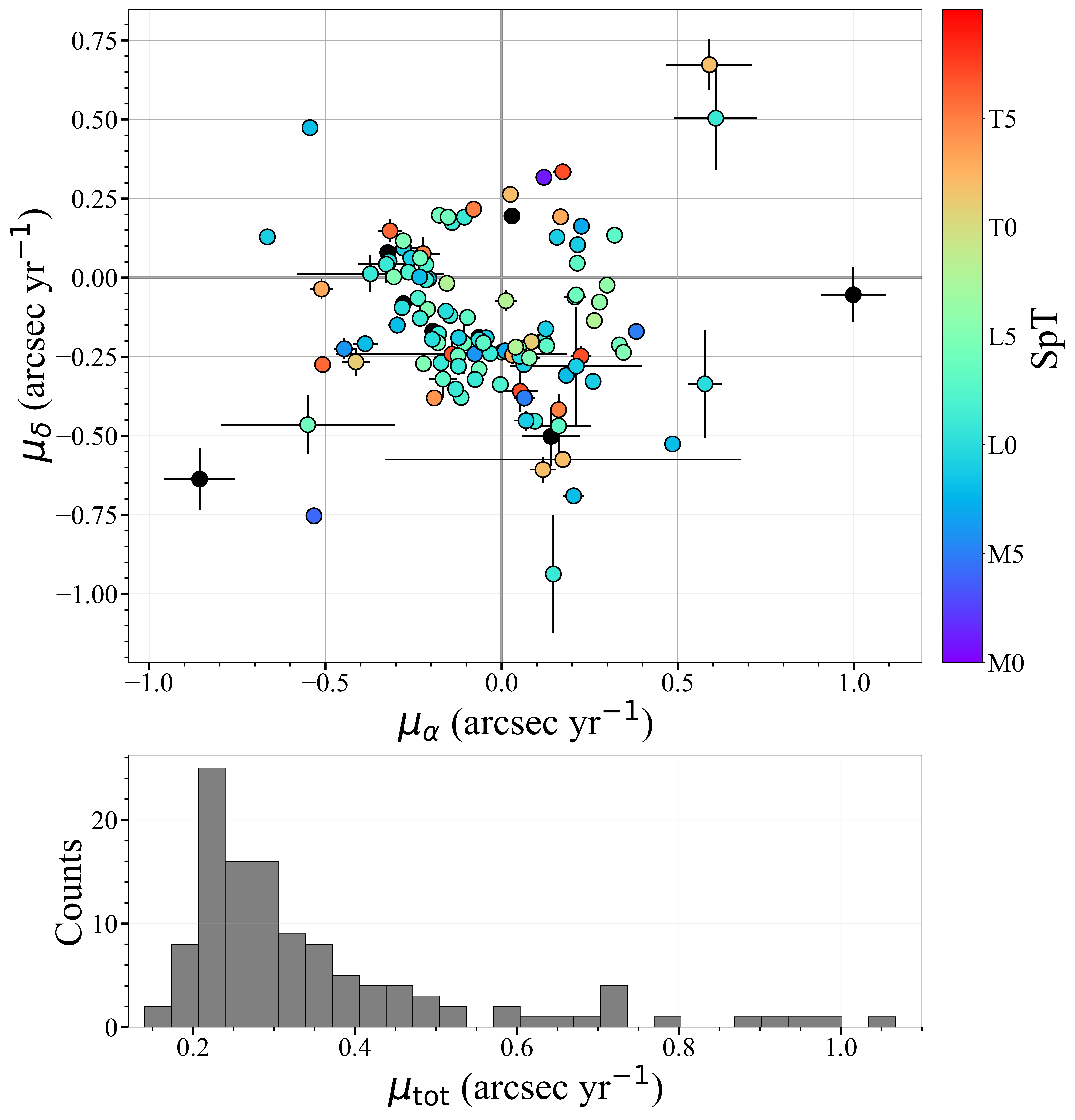}
        \caption{The top plot compares $\mu_{\alpha}$ to $\mu_{\delta}$ (arcsec yr$^{-1}$), with symbol color encoding photometric classification. The black points with associated numbers are subdwarfs described in \S\ref{cheetoo}. The bottom plot shows the distribution of the total proper motion of the \texttt{SMDET} sample. Note that objects with calculated proper motions were utilized in the analysis, while all other objects relied on proper motion measurements from CatWISE2020.}
        \label{Figure 8}
\end{figure}

The proper motion distribution and orientations in our sample are shown in Figure \ref{Figure 8}. This figure highlights that our sample has a minimum detection limit for $\mu_{tot}$ (total proper motion) of about 0.2 (arcsec yr$^{-1}$). This is not because \texttt{SMDET} fails to identify low proper motion objects, rather \texttt{SDMET} was trained with solely high proper motion examples thus assigning low proper motion objects a lower scores. Low proper motion candidates are overwhelmed by false positives that likely lead to them being overlooked during human verification. As a result, these low proper motion candidates are often overshadowed by false positives and may be overlooked during human verification. Therefore, this minimum detection limit reflects both a human bias and an inherent feature of the \texttt{SMDET} training process. To achieve a lower minimum proper motion detection limit, a new training set would need to be developed and run through the neural network, which future surveys should aim to accomplish. 

The distribution of photometric classifications is shown in Figure \ref{Figure 9}. This figure does not include the 3 subdwarf candidates discussed in \S\ref{cheetoo}. We anticipated a bias toward later spectral types due to this study focusing on colder, fainter sources. However, our distribution suggests the opposite trend. This discrepancy can be attributed to the sensitivity limit of the photometric surveys, which restricts the volume in which each spectral type can be detected. The brighter, earlier-type dwarfs are simply detected out to a larger volume. Discovering later-type brown dwarfs at increased distances will need next-generation infrared surveys, such as the \textit{Near-Earth Object Surveyor} (\citealt{Mainzer et al.(2023)}), Nancy Grace Roman Space Telescope (\citealt{Kasdin et al.(2020)}), Euclid (\citealt{Racca et al.(2016)}), and others. 

\section{Conclusion}\label{conclusion}
We report the discovery of 118 new ultracool dwarf candidates. We combined photometry and astrometry from multiple catalogs to derive estimates for spectral type, distance, and tangential velocity. Our sample includes 28 M, 64 L, and 18 T dwarfs, and we verified the spectral types of two T dwarf candidates using near-infrared spectroscopy. We also identified one new T subdwarf and two new extreme T subdwarf candidates, based on their similar colors to previously confirmed subdwarfs; and two low-probability candidate young ultracool dwarfs

\section{Acknowledgments}
Backyard Worlds research was supported by NASA grant 2017-ADAP17-0067 and by the NSF under grants AST-2007068, AST-2009177, and AST-2009136. This work makes use of data products from WISE/NEOWISE, which is a joint project of UCLA and JPL/Caltech, funded by NASA. The CatWISE effort was led by the Jet Propulsion Laboratory, California Institute of Technology, with funding from NASA’s Astrophysics Data Analysis Program. The observations obtained as part of the VISTA Hemisphere Survey, ESO Progam, were funded by the grant 179.A-2010 (PI: McMahon). The UHS is a partnership between the UK STFC, The University of Hawaii, The University of Arizona, Lockheed Martin, and NASA.

The PanSTARRS2 Surveys (PS2) and the PS2 public science archive have been made possible through contributions by the Institute for Astronomy, the University of Hawaii, the PanSTARRS Project Office, the Max-Planck Society and its participating institutes, the Max Planck Institute for Astronomy, Heidelberg and the Max Planck Institute for Extraterrestrial Physics, Garching, The Johns Hopkins University, Durham University, the University of Edinburgh, the Queen's University Belfast, the Harvard-Smithsonian Center for Astrophysics, the Las Cumbres Observatory Global Telescope Network Incorporated, the National Central University of Taiwan, the Space Telescope Science Institute, the National Aeronautics and Space Administration under Grant No. NNX08AR22G issued through the Planetary Science Division of the NASA Science Mission Directorate, the National Science Foundation Grant No. AST-1238877, the University of Maryland, Eotvos Lorand University (ELTE), the Los Alamos National Laboratory, and the Gordon and Betty Moore Foundation. 

This paper uses observations obtained at the Hale Telescope, at Palomar Observatory, which is part of a continuing collaboration between the California Institute of Technology, NASA/JPL, Yale University, and the National Astronomical Observatories of China.

This research has made use of the VizieR catalog access tool, CDS,
Strasbourg, France (DOI : 10.26093/cds/vizier). The original description of the VizieR service was published in 2000, A$\&$AS 143, 23. This work has made use of data from the European Space Agency (ESA) mission {\it Gaia} (\url{https://www.cosmos.esa.int/Gaia}), processed by the {\it Gaia} Data Processing and Analysis Consortium (DPAC, \url{https://www.cosmos.esa.int/web/Gaia/dpac/consortium}). Funding for the DPAC has been provided by national institutions, in particular, the institutions participating in the {\it Gaia} Multilateral Agreement.

\software{
WiseView (\citealt{Caselden et al.(2018)}),
SpeXTool \citep{Cushing et al.(2004)},
Matplotlib \citep{Hunter(2007)},
NumPy \citep{van der Walt et al.(2011)}
}


\end{document}